%% file: main.tex
\documentclass{Interspeech2024}
\usepackage{bm}

\interspeechcameraready

\title{DeWinder: Single-Channel Wind Noise Reduction using Ultrasound Sensing}

\name[affiliation={1*}]{Kuang}{Yuan}
\name[affiliation={1*}]{Shuo}{Han}
\name[affiliation={1}]{Swarun}{Kumar}
\name[affiliation={1}]{Bhiksha}{Raj}

\address{
  $^1$Carnegie Mellon University, United States
}
\email{kuangy@cmu.edu, shuohan@andrew.cmu.edu, swarun@cmu.edu, bhiksha@cs.cmu.edu}


\newcommand{\sname}{\emph{DeWinder}}

\begin{document}

\maketitle
\def\thefootnote{*}\footnotetext{Equal Contribution}\def\thefootnote{\arabic{footnote}}

\begin{abstract}
The quality of audio recordings in outdoor environments is often degraded by the presence of wind. Mitigating the impact of wind noise on the perceptual quality of single-channel speech remains a significant challenge due to its non-stationary characteristics. Prior work in noise suppression treats wind noise as a general background noise without explicit modeling of its characteristics. In this paper, we leverage ultrasound as an auxiliary modality to explicitly sense the airflow and characterize the wind noise. We propose a multi-modal deep-learning framework to fuse the ultrasonic Doppler features and speech signals for wind noise reduction. Our results show that \sname\ can significantly improve the noise reduction capabilities of state-of-the-art speech enhancement models.

\end{abstract}

\input{Introduction}

\input{dataset}
\input{model}

\input{experiment}

\input{conclusion}
\input{acknowledgement}
\bibliographystyle{IEEEtran}
\bibliography{mybib}

\end{document}

%% file: Introduction.tex
\vspace{-0.01in}
\section{Introduction}

Wind noise is a major noise source that degrades the audio recording quality in various environments with the presence of airflow.
Different from other noise sources, wind noise is generated by turbulent airflow hitting the microphone membrane instead of propagating acoustic waves~\cite{turbulent}.
Due to the non-stationary nature of turbulence, suppressing wind noise in audio remains an open challenge. 
A microphone windshield is used to reduce the excessive pressure from wind in professional use cases. However, such hardware solutions are not suitable for tiny microphones on embedded devices like smartphones.

To mitigate the interference of wind noise, a series of prior works~\cite{multich-windnoise1, multich-windnoise2, multich-windnoise3, multich-windnoise4, multich-windnoise45} propose wind noise estimation and reduction algorithms relying on multi-channel microphone arrays, leveraging the assumption that wind noise is spatially uncorrelated. In contrast, for single-channel audio, spectral subtraction~\cite{specsubtract, specsubtract2, specsubract3} and filtering~\cite{postfiltering, filtering2, emdfiltering} algorithms are proposed to reduce wind noise. However, such techniques are based on the stationary assumption when estimating the noise spectral distribution, the performance of which can drop significantly under real-world non-stationary wind noise. Recent speech enhancement techniques~\cite{dccrn, demucs, hao2020fullsubnet, RemixIT, mfnet, dctcrn} based on Deep Neural Networks~(DNN) have shown promise in removing undesire noise in audio. However, existing speech enhancement models only treat wind noise as a general background noise without explicitly modeling its characteristics, which may cause sub-optimal performance, especially in strong wind environments with low Signal-to-Noise Ratio~(SNR).

Instead of relying on pre-estimated noise distribution, we propose incorporating a new modality to sense and characterize the real-time airflow profile, and further enable more informative wind noise reduction. Specifically, for the first time, we propose to utilize ultrasound as a complementary modality to gather information about wind noise. Our design is based on a key observation: the ambient airflow not only introduces wind noise when hitting the microphone but also shapes the propagation of other acoustic signals in the air. Intuitively, if an acoustic wave is in the same direction as the airflow, it will travel faster than its original speed, such that its frequency becomes higher at the receiver because of the Doppler effect~\cite{doppler}. More generally, the turbulent airflow that induces wind noise contains many unsteady vortexes moving toward different directions, which shapes the frequency of the acoustic signals in a more unstructured way. We propose to use high-frequency ultrasound signals to sense and characterize the airflow, by capturing such frequency differences caused by the Doppler effect with finer granularity than audible signals, without inducing any audible disturbances.

\begin{figure*}
    \begin{minipage}{0.30\textwidth}
\centering
        \vspace{-0.01in}
        \includegraphics[width=\textwidth]{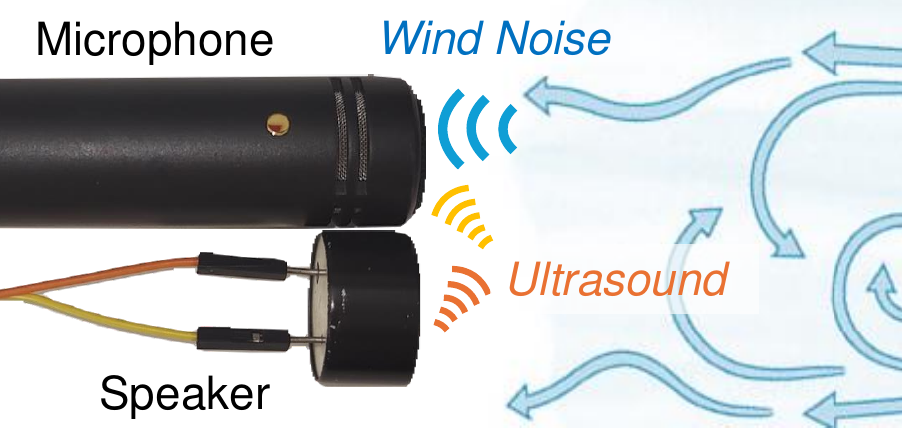}
    \caption{\sname\ uses ultrasound to sense and reduce wind noise.}
    \label{fig:setup}
\end{minipage}
\hspace{0.05in}
\begin{minipage}{0.33\textwidth}
    \centering
        \vspace{-0.2in}
        \includegraphics[width=\textwidth]{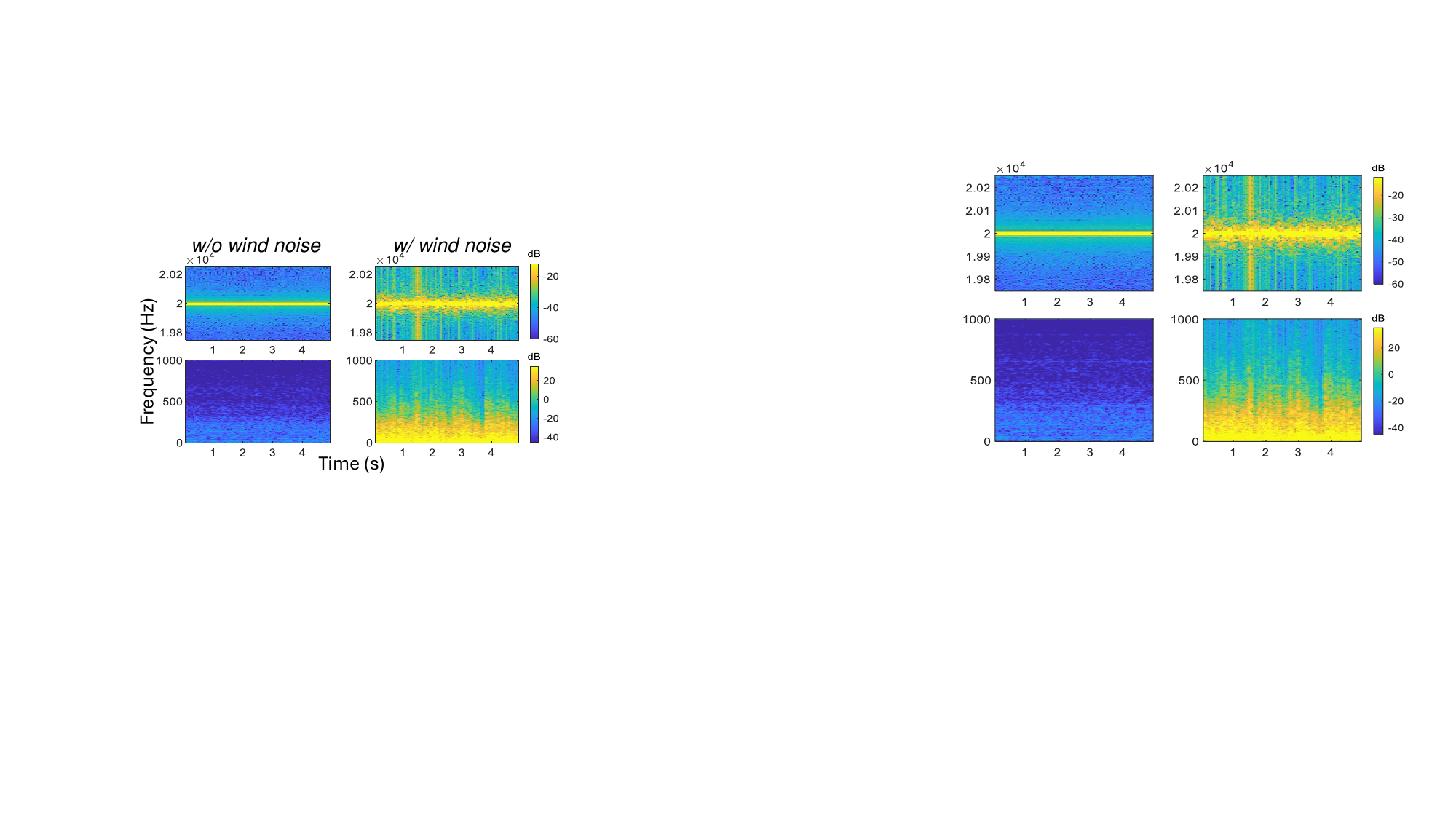}
        \vspace{-0.2in}
    \caption{The airflow induces wind noise, while shaping the ultrasound transmission. }
    \label{fig:enter-label}
\end{minipage}
\hspace{0.03in}
\begin{minipage}{0.35\textwidth}
    \centering
        \vspace{-0.08in}
        \includegraphics[width=\textwidth]{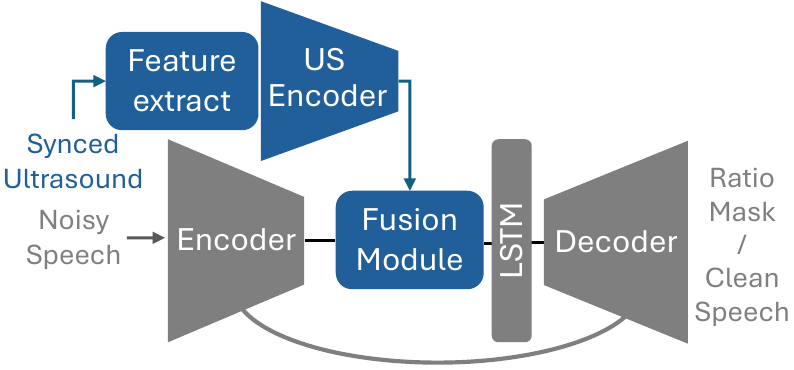}
    \caption{Modular Design that can be adapted to existing speech enhancement models.}
    \label{fig:arch}
\end{minipage}
\end{figure*}

 Specifically, we implement the idea of \sname\ by using an ultrasound speaker co-located with the microphone.  The ultrasound speaker transmits a tone signal at around 20~kHz, which is not audible but can still be captured by commodity microphones sampling at 44.1 kHz. As illustrated in Fig.~\ref{fig:setup}, the signal transmitted from the speaker (Orange) would be shaped by airflow near the device before arriving at the microphone. The microphone would simultaneously capture the wind noise (Blue), as well as the ultrasound signal (Yellow) carrying information about the real-time airflow profile. 
 We note that our proposed hardware setup can be easily extended to off-the-shelf IoT devices with co-located speakers and microphones~\cite{sun2021ultrase, earphonetrack}, such as smartphones (located at the bottom of the screen).

To implement \sname, we design a modular framework that can be adapted to different existing speech enhancement DNN models to improve the wind noise reduction performance through speech-ultrasound multi-modal fusion. As shown in Fig.~\ref{fig:arch}, we first develop an ultrasound feature extraction pipeline consisting of demodulation and multi-step filtering, to convert ultrasound signal into baseband waveforms that focus on the Doppler effects induced by airflow. We demonstrate the performance of \sname\ by adapting the architecture of two state-of-the-art speech enhancement models: DEMUCS~\cite{demucs} and DCCRN~\cite{dccrn}. Both of the networks have a convolutional encoder-decoder architecture with a unidirectional LSTM in-between for sequence modeling. For both of the architectures, we design another convolution encoder specialized for ultrasound encoding, and fuse the extracted ultrasound embedding with the speech embedding before the LSTM through a customized fusion module. While the two models process the speech in different domains (DEMUCS - waveform domain, DCCRN - spectrogram domain),  our results show that the \sname's fusion framework can improve the wind noise reduction performance in both domains, especially in low-SNR conditions.

%% file: dataset.tex
\section{Dataset Collection and Processing}
\label{sec:dataset}

\subsection{Dataset Collection}

As visualized in Fig.~\ref{fig:setup}, the ultrasound speaker is placed co-located with the microphone, and towards the same direction. Specifically, we place two ultrasound speakers at the left and right side of the Rode M5 microphone, transmitting tones at 20 and 21~kHz respectively, to ensure the sensitivity of the system is symmetric to the wind from different azimuth directions (We also present the evaluation results with only a single speaker in Sec.~\ref{sec:experiment}). The microphone samples at 44.1~kHz and is connected to a laptop through an RME Babyface Pro sound card~.

To generate consistently powerful enough airflow that can induce a significant amount of wind noise to the microphone, we use two types of high-velocity fans to generate wind. We collect the dataset in three different indoor environments to enrich the diversity of turbulence. We further place the fan along four different directions relative to the microphone in each environment at around 2-5~m away from the microphone, and allow it blow wind towards the microphone. The average wind speed in front of the microphone membrane is 1-2.5 m/s across different environments and distances, as measured by an anemometer. We collect the dataset of wind noise along with ultrasound signal transmissions for 4.2 hours, and further synthesize a 10.3-hour dataset by mixing it with clean speech utterances (detailed in Sec.~\ref{subsec:syn}).

\subsection{Preprocessing and Feature Extraction}

During data collection, both wind noise (mostly $<$1~kHz) and ultrasound signals (around 20 and 21~kHz) are captured by the microphone. To obtain the noise independently, we first apply a low-pass filter to extract the signal under 1.2~kHz. We further apply a high-pass filter with a cutoff frequency of 20~Hz to suppress the inaudible low-frequency artifacts caused by the wind. Finally, we resample the noise signal to 16~kHz to let it match with the speech dataset and meet the input requirements of most speech enhancement models~\cite{demucs, dccrn, RemixIT}.

For ultrasound processing, our primary goal is to convert high-frequency signals into representations that can be handled by common CNN encoders, as well as capture the frequency differences induced by the Doppler effect. We choose to down-convert the ultrasound signals into baseband waveforms by mixing them with sine and cosine waves at the corresponding center frequencies (20 or 21~kHz) and applying low-pass filtering. Through further resampling to 16~kHz, such baseband waveforms can be processed by the encoders of speech enhancement models and retain the characteristics of sample-level synchronization with noise. As a normal airflow with a wind speed of less than 8~m/s will only introduce a Doppler shift of less than 500~Hz to the 20~kHz ultrasound, we choose the low-pass filter with a cutoff frequency of 500~Hz, to let the signal focus on the features produced by the Doppler effect, as well as remove the high-frequency components introduced by mixing~\cite{downconvert}. We further apply a highpass filter at a 10~Hz cutoff frequency to mitigate the signals reflected from nearby static objects. Through the above processing, each ultrasound signal is converted into two channels of baseband waveforms (mixed through sine and cosine waves). Thus, along with a single-channel wind noise, we obtain four channels of sample-level synchronized waveforms that help to characterize the wind noise.

%% file: model.tex
\section{Model}

Instead of building an architecture from scratch, we design \sname\ as a modular framework that can be adapted to different existing speech enhancement models. We demonstrate \sname\ on DEMUCS and DCCRN. Both models have a convolutional encoder-decoder architecture with U-Net skip connections~\cite{ronneberger2015unet} and an LSTM in between. As shown in Fig.~\ref{fig:arch}, we design another branch of an ultrasound encoder in parallel with the speech encoder, that processes the multi-channel waveforms (extracted from ultrasound, Sec.~\ref{sec:dataset}) into embeddings that characterize wind noise. 
We now note a key difference between the embeddings generated by DCCRN and DEMUCS that informs \sname's fusion design. DEMUCS captures a speech embedding that is designed to generate clean speech output. In contrast, DCCRN's embedding captures the noise rather than the speech, to estimate a ratio mask output that helps separate noise from speech. Given that the two model embeddings  have entirely different semantics, the fusion module \sname\ employs prior to LSTM for DEMUCS and DCCRN are customized accordingly. We detail \sname's ultrasound encoder and fusion module for DEMUCS and DCCRN respectively below. 


\subsection{Dewinder - DEMUCS}

DEMUCS~\cite{demucs} is a speech enhancement model that operates on the waveform domain. DEMUCS extracts the speech embeddings through an encoder from noisy speech and seeks to reconstruct clean speech at the decoder using the embeddings. For the ultrasound encoder, we follow a similar architectural design approach as the original speech encoder in DEMUCS. As the waveforms extracted from ultrasound have a narrower frequency range ($<$500~Hz) than speech, we reduce the number of CNN layers in the encoder to three to mitigate overfitting. Additionally, we set the base hidden channel size $H$ to 24 instead of the default setup of 48, and the convolutional kernel size $K$ to 10. We ensure that both the speech and ultrasound encoders maintain temporal alignment during convolutions, enabling coherent integration of the two modalities. We note that the input of the ultrasound encoder is the four-channel waveforms extracted from ultrasound, instead of a single channel.

\begin{figure}
    \centering
    \includegraphics[width=0.26\textwidth]{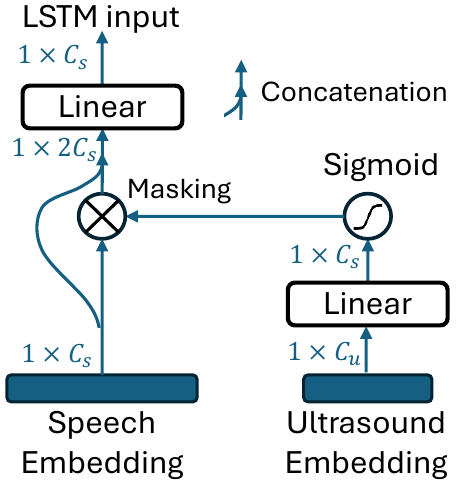}
    \vspace{-0.05in}
    \caption{Fusion Module for DEMUCS based on Masking}
    \label{fig:fusin-demucs}
    \vspace{-0.15in}
\end{figure}

The encoder of DEMUCS seeks to extract embeddings representing clean speech from noisy speech, while inevitably retaining noise information in the embedding for low-SNR inputs. Thus, to further mitigate the wind noise corruption, we design a fusion module based on masking that leverages the ultrasound embedding to \textit{filter} out the wind noise information in the speech embedding, before feeding it into the LSTM.
As illustrated in Fig.~\ref{fig:fusin-demucs}, we first apply a linear layer to the ultrasound embedding to let its dimension match with speech embedding. We then apply an element-wise Sigmoid activation and multiply it with the speech embedding, which is essentially an embedding mask to suppress the wind noise information. Mathematically, 
\begin{equation}
    \bm{X}_s' = \bm{X}_s\cdot\sigma(\bm{X}_u\bm{W}^T+\bm{b})
\end{equation}

where $\bm{X}_s, \bm{X}_u$ are the speech and ultrasound embedding respectively with dimensions illustrated in the figure. $\bm{W}$ and $\bm{b}$ are the weights and bias of the linear layer, and $\bm{X}_s'$ is the denoised speech embedding. We further concatenate the denoised embedding with the original speech embedding and apply another linear layer to project the dimension back to the original input dimension of the LSTM.

\subsection{Dewinder - DCCRN}
Deep Complex Convolution Recurrent Network (DCCRN) is a speech enhancement model that operates on the Time-Frequency (TF) Domain. 
The encoder extracts the noise information from the complex-value spectrogram into embeddings. The LSTM and decoder process the embeddings and estimate a Complex Ratio Mask (CRM)~\cite{crm}, which can be applied to the original spectrogram for noise reduction. 

Similarly, we design an encoder that extracts the wind noise information from ultrasound inputs by adapting the design of its original encoder. In the DCCRN-CL configuration, six layers of CNN have numbers of channels of \{16,32,64,128,256,256\}, with kernel size and stride of (5,2) and (2,1) respectively. To reduce the parameter size, we choose a five-layer design with fewer output channels of \{16,32,64,128,128\} to mitigate overfitting. To ensure time-window synchronization of the extracted embedding with the embedding extracted from speech, we modify the stride of the first layer of ultrasound encoder into (4,1). 

In the fusion module, as the embeddings from two encoders are both designed to represent the wind noise information, we choose to combine the information from the two modalities by concatenating the embeddings. Specifically, the complex-value embedding from the speech encoder can be represented by $\bm{X}=\bm{X}_r+j\bm{X}_i$, where $\bm{X}_r, \bm{X}_i\in\mathbb{R}^{C_s\times F\times T}$, and ultrasound embedding can be represented by $\bm{Y}=\bm{Y}_r+j\bm{Y}_i$, where $\bm{Y}_r, \bm{Y}_i\in\mathbb{R}^{C_u\times F\times T}$ and $j=\sqrt{-1}$. $C_s, C_u$ are the channel dimensions of speech embedding and ultrasound embeddings respectively. $F, T$ are the frequency and time dimensions. We concatenate the real part and imaginary part of the two embeddings along the channel axis, before passing these into the complex LSTM.

As the first paper exploring using ultrasound for wind noise reduction, \sname\ shows the modular design by adapting two speech enhancement models, DCCRN and DEMUCS. The two models process the speech in different domains as well as have different forms of decoder outputs. We envision our designs have a great potential to be generalized to other speech enhancement models that share similar structures.



%% file: experiment.tex
\section{Experiment}
\label{sec:experiment}
\subsection{Datasets Synthesization}
\label{subsec:syn}
In our experimental setup, we train and evaluate the proposed methods and baseline models on synthesized noisy datasets. We collect 4.2 hours of wind noise dataset along with ultrasound transmission. We randomly split the audio into 3.36 hours and 0.84 hours respectively for training and validation, ensuring no overlapping. In each set, we extract 5-second audio segments in a sliding window manner with a hop size of 2 seconds to augment the size of the dataset. We then randomly select 7420 clean utterances (each with 5 seconds) in total from LibriSpeech dataset~\cite{librispeech} and mix them with wind noise at random low SNR between -40 dB and -20 dB. We also ensure that there is no overlap in speaker identities between the clean utterances used for training and validation.  During mixing, we fix the power of wind noise to ensure it still physically matches with ultrasound, and tune the power of speech signal to generate mixed speech at different SNR. We note that the SNR values we used for synthesizing are lower than the ones used in other prior speech enhancement works~\cite{dccrn, demucs}, mainly because we consider wind noise as the only noise source. We choose the range of SNR at -40 to -20 dB as it perceptually matches the audio recordings quality in real-world outdoor windy environments during our initial subjective testing.

Finally, we get a training and validation set with 5954 and 1466 5-second audio respectively (8.27 and 2.03 hours in total). Each sample includes a speech signal corrupted by wind noise with two ultrasound signals, and a corresponding clean speech signal as the ground truth. For the testing set, we randomly shuffle the wind noise audio segments and clean speech utterances used in the validation set, as well as re-generate SNR values in the range, to create a new set of mixed audio segments.


\subsection{Training Setup and Baselines}

We first train the original architecture of DCCRN and DEMUCS as our baseline. Specifically, we choose the configuration that with the best performance reported in the paper, namely DCCRN-CL and DEMUCS (H=48, S=4, U=4). we use the causal setup in both models where the LSTM is unidirectional. We use the AdamW optimizer~\cite{adamw} with a learning rate of 3e-4, a momentum of $\beta_1=0.9$, a denominator momentum of $\beta_2=0.999$, and a weight decay of 1e-3. The models are selected by early stopping. Model training are testing are performed on a NVIDIA - GeForce RTX 3090 Graphics Card.

For DEMUCS training, we use the loss presented in the paper, which is the sum of the waveform L1 loss and the multi-resolution STFT loss. Given clean signal $\bm{y}$ and estimated signal $\hat{\bm{y}}$ from the model, the loss function can be represented by:
$$
L(\bm{y}, \hat{\bm{y}}) = \frac{1}{T}||\bm{y}-\hat{\bm{y}}||_1 + L_{stft}(\bm{y}, \hat{\bm{y}})
$$
where T is the number of samples in the waveform, and we refer the definition of STFT loss in DEMUCS paper~\cite{demucs}.

For DCCRN training, instead of using the SI-SNR loss~\cite{luo2018tasnet} alone defined in the paper, we also incorporate the STFT loss presented in DEMUCS to improve the overall perceptual audio quality. The loss can be represented by:
$$
L(\bm{y}, \hat{\bm{y}}) = \lambda_1 10\log_{10}(\frac{||\bm{y}_{target}||_2^2}{||\hat{\bm{y}}-\bm{y}||_2^2}) + \lambda_2 L_{stft}(\bm{y}, \hat{\bm{y}})
$$

where $\bm{y}_{target} = <\hat{\bm{y}},\bm{y}>\cdot \bm{y} / ||\bm{y}||_2^2$. During our training, $\lambda_1$ is set to -0.2 and $\lambda_2$ is set 1.

Once the baseline model is converged, we add the module of \sname\ including the ultrasound encoder and the fusion module to the network. For the original encoder, decoder, and LSTM in the model, we load the weight that was pre-trained in the baseline model as initialization. We note that the first LSTM layer in DCCRN is re-initialized as the input dimension is modified. We train the two-stream architecture of \sname\ using the same optimizer setup as the baseline training.

\subsection{Evaluation Results}

\begin{table}
\centering
\begin{tabular}{lccc}
\toprule
& SI-SDR & PESQ & STOI \\
\midrule
DCCRN (3.7~M) & 2.685 & 2.265 & 0.653 \\
\sname\ - Original Loss & 3.581 & 2.374 & 0.671 \\
\sname\ - Single & 3.841 & 2.470 & 0.696 \\
\sname\ (4.2~M) & \textbf{3.871} & \textbf{2.480} & \textbf{0.700} \\
\midrule
DEMUCS (18.9~M) & 6.632 & 2.776 & 0.812 \\
\sname\ - Concat Fusion  & 6.057 & 2.805 & 0.820 \\
\sname\ - Single & 6.902 & 2.855 & 0.826 \\
\sname\ (21.7~M) & \textbf{6.932} & \textbf{2.861} & \textbf{0.827} \\

\bottomrule
\end{tabular}
\caption{Performance of \sname\ and ablation study}
\vspace{-0.2in}
\label{tab:main-result}
\end{table}

We evaluate the performance of \sname\ of the adaptations on DCCRN and DEMUS separately, as the two baseline models have different levels of parameter sizes (3.7~M and 18.9~M). We use the evaluation metrics including Scale-Invariant Signal-to-Distortion Ratio (SI-SDR)~\cite{sisdr}, Perceptual Evaluation of Speech Quality (PESQ)~\cite{pesq} (from 0.5 to 4.5), and Short-Time Objective Intelligibility (STOI)~\cite{stoi} (from 0 to 1). 

Table~\ref{tab:main-result} shows the performance of evaluation on our overall testing set with SNR values in the range of -40 to -20 dB. Besides the baseline models of DCCRN and DEMUCS and the \sname's complete setup, we also show the performance of \sname\ setup but with the original SI-SNR loss used in the DCCRN paper~(\sname\ - Original Loss). We also modify the input channel number of the ultrasound encoder and let it take only the waveforms from a single ultrasound speaker (20 or 21~kHz). We average the model performance across two frequencies (\sname\ - Single). For the evaluation based on DEMUCS,  we also evaluated the model that employs concatenation as the fusion module (\sname\ - Concat Fusion) to compare with our proposed masking-based fusion.

The results show that adding \sname\ to the baseline models can significantly improve the wind noise reduction performance. Surprisingly, we observe that \sname\ based on only single ultrasound speaker can achieve almost the same level of performance compared to the full two-speaker setup, which demonstrates the potential capability of \sname\ to be deployed on the current hardware setup on off-she-shelf devices such as smartphones, with only a single side of speaker. Meanwhile, we report the total parameter size of the baseline models and \sname\ in the table. Our design does not need to introduce a significant amount of parameters to the original network, since both the ultrasound encoder and fusion module are lightweight, and thus would not degrade the real-time interference capability of the model.

We also observe the concatenation-based fusion in DEMUCS (\sname\ - Concat Fusion) cannot significantly improve the performance compared to the baseline, and even obtains a lower SI-SDR value. Such a result demonstrates the necessity of our masking-based fusion module which considers the semantic meaning of the embeddings.

We further present the evaluation result of the performance improvement at different SNRs. We synthesize the testing sets using fixed pairs of wind noise and speech, but repeat the mixing at different SNR values. We evaluate the PESQ and STOI improvements at SNRs from -35 to -10 dB compared with baseline models. We observe similar trends in both DCCRN and DEMUCS. The performance improves significantly in low-SNR scenarios under -25 dB, while still outperforming the baseline in higher SNR cases. We observe that the adaptation on DCCRN achieves a higher value of performance improvements than the adaptation on DEMUCS. We attribute this to the larger parameter size of DEMUCS, and better capability to reduce non-stationary noise of the waveform-domain models. Thus the baseline DEMUCS model can achieve better performance in wind noise reduction and leave less room for improvement.

\begin{figure}
    \centering
    \vspace{-0.1in}
    \includegraphics[width=0.37\textwidth]{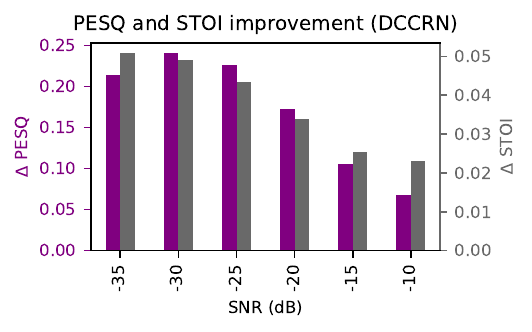}
    \vspace{-0.1in}
    \label{fig:enter-label}
    \includegraphics[width=0.37\textwidth]{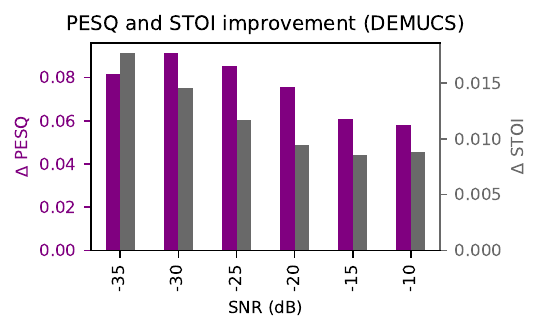}
    \vspace{-0.15in}
    \caption{Performance improvement at different SNRs }
   \vspace{-0.15in}
    \label{fig:s}
\end{figure}

%% file: conclusion.tex
\section{Conclusion}
In this work, we present \sname, which for the first time, utilizes ultrasound as a complementary modality to sense the wind noise and perform noise reduction. We design a modular framework that can be adapted to different speech enhancement models without introducing significant computational overhead. We collect a wind noise dataset along with ultrasound transmissions and demonstrate \sname\ can significantly improve the wind noise reduction capability of two state-of-the-art speech enhancement models: DEMUCS and DCCRN. We leave the design of more complex transmitted ultrasound signals, and exploring other multi-modal fusion mechanisms for further work.

%% file: acknowledgement.tex
\section{Acknowledgement}
We acknowledge support from the NSF (2106921, 2030154, 2007786, 1942902, 2111751), ONR, AFRETEC, MFI, CISCO, Safety21 and CyLab-Enterprise. We thank the anonymous reviewers for their constructive feedback. 